\documentclass[useAMS,usegraphicx, usenatbib]{mn2e}
\usepackage{aas_macros}

\usepackage{amsmath}
\usepackage{amssymb}
\usepackage{graphicx}
\usepackage{latexsym}
\usepackage{graphicx}
\usepackage{epsfig}
\usepackage{dcolumn}
\usepackage{bm}
\usepackage{natbib}
\usepackage{setspace}
 \usepackage{psfrag}
\usepackage{color}
\usepackage[usenames,dvipsnames]{xcolor}

\begin{document}
\title{Constraints on Warm Dark Matter models from high-redshift long gamma-ray bursts}

\author[R. S. de Souza,  A.  Mesinger,  A. Ferrara,   Z. Haiman, R. Perna, N. Yoshida]
{R. S. de Souza  $^{1}$\thanks{e-mail: rafael@kasi.re.kr (RSS)}, A.  Mesinger $^{2}$,  A. Ferrara $^{2}$,   Z. Haiman $^{3}$, R. Perna $^{4}$, N. Yoshida $^{5}$
\\
$^{1}$Korea Astronomy \& Space Science Institute, Daejeon 305-348, Korea\\
$^{2}$Scuola Normale Superiore, Piazza dei Cavalieri 7, 56126 Pisa, Italy\\
$^{3}$Department of Astronomy, Columbia University, 550 West 120th Street, New York, NY 10027, USA\\
$^{4}$JILA and Department of Astrophysical and Planetary Science, University of Colorado at Boulder, 440 UCB, Boulder, CO, 80309, USA\\
$^{5}$Department of Physics, University of Tokyo, Tokyo 113-0033, Japan\\
}

 \date{Accepted -- Received  --}

\pagerange{\pageref{firstpage}--\pageref{lastpage}} \pubyear{2013}

\voffset-.6in

\maketitle
\label{firstpage}

\begin{abstract}
Structures in Warm Dark Matter (WDM) models are exponentially suppressed below a certain scale, characterized by the dark matter particle mass, $m_{\rm x}$.  Since structures form hierarchically, the presence of collapsed objects at high-redshifts can set strong lower limits on $m_{\rm x}$. We place robust constraints on $m_{\rm x}$ using recent results from the {\it Swift} database of high-redshift gamma-ray bursts (GRBs).  We parameterize the redshift evolution   of the ratio between the cosmic GRB rate and star formation rate (SFR) as $\propto (1+z)^\alpha$, thereby allowing astrophysical uncertainties to partially mimic the cosmological suppression of structures in WDM models.  Using a maximum likelihood estimator on two different $z>4$ GRB subsamples (including two  bursts  at $z>8$), we constrain $m_{\rm x} \gtrsim 1.6$-1.8 keV at 95\% CL, when marginalized over a flat prior in $\alpha$.  We further estimate that 5 years of a SVOM-like  mission would tighten these constraints to $m_{\rm x} \gtrsim 2.3 $ keV.    Our results show that GRBs are a powerful probe of high-redshift structures, providing robust and competitive constraints on $m_{\rm x}$.

\end{abstract}

\section{Introduction}

The current concordance cosmology, in which structure formation proceeds in a hierarchal manner driven by pressureless cold dark matter (CDM), has been remarkably successful in  explaining  the  observed properties of large-scale structures in the Universe  \citep[e.g.,][]{Tegmark2006,Benson2010} and the cosmic microwave background (CMB) \citep[e.g.,][]{Komatsu2011}. Such observables  probe  scales in the range  $\sim$ 1 Gpc down to $\sim$ 10 Mpc. 
On smaller scales, $\lesssim$ 1 Mpc, there are still  some  discrepancies between standard $\Lambda$CDM and observations \citep[e.g., ][]{Menci2012}. For instance,  N-body simulations predict more satellite galaxies than are observed both around our galaxy (the so-called ``missing satellite problem''; \citealt[e.g.,][]{moore1999,Klypin1999}), and in the field as recently noted by the ALFALFA survey \citep[e.g.,][]{Papastergis2011,Ferrero2012}.  Furthermore, simulations of the most massive Galactic CDM subhaloes are too centrally condensed  to be consistent with the kinematic data of  the bright Milky Way satellites \citep[e.g.,][]{Boylan2011}. Moreover, observations of small galaxies  show that their central   density profile is  shallower than predicted by CDM N-body simulations  \citep[e.g.,][]{moore1994,deblok2001,Donato2009,Maccio2012,Governato2012}.

Baryonic feedback is a popular prescription for resolving such discrepancies.
Feedback  caused by supernovae (SNe) explosions and heating due to the UV background may suppress the baryonic content of low-mass haloes  \citep[e.g.,][]{Governato2007, Mashchenko2008, Busha10, SM2013a}, and make their inner  density profile shallower \citep[e.g.,][]{desouza2010,desouza2011a}.  However, accurately matching observations is still difficult even when tuning feedback recipes (e.g. \citealt{B-K12}).

An alternative explanation might be found if dark matter (DM) consisted of lower mass ($\sim$keV) particles and thus was ``warm'' (WDM; e.g.,  \citealt{Bode2001, Khlopov2008, deVega2012,VSS2012,Kang2013,Destri2013,Kamada2013}).  The resulting effective pressure and free-streaming would decrease structure on small-scales, though again fine-tuning might be required to fully match all the observations (e.g. \citealt{Boylan2011, Maccio2012, Borriello12}).

The most powerful testbed for these scenarios is the high-redshift Universe.  Structure formation in WDM models (or in any cosmological model with an equivalent power-spectrum cut-off) is exponentially suppressed on small-scales \citep[e.g.,][]{Schneider2012,Schneider2013}.  Since structures form hierarchically, these small halos are expected to host the first galaxies.  If indeed dark matter were sufficiently ``warm'', the high-redshift Universe would be empty.  Therefore, the mere presence of a galaxy at high-redshift can set strong lower limits on the WDM particle mass.

Due to their high luminosity,  gamma-ray bursts (GRBs) constitute a remarkable tool to probe the high-$z$ Universe and small-scale structures. They provide a glimpse of the first generations of stars  \citep[e.g.,][]{rafael2011,deSouza2012}, as well as provide constraints on primordial non-Gaussianity \citep{Maio2012}.  As pointed out by \citet{Mesinger2005},  the detection of a single GRB at $z > 10$ would provide very strong constraints on WDM models.

Here we extend the work of \citet{Mesinger2005} by {\it presenting robust lower limits on WDM particle masses, using the latest \textit{Swift} GRB data.}
The current data, including many redshift measurements,   allows us to perform an  improved statistical analysis by directly comparing the distribution of bursts in various models as a function of  redshift.  Furthermore, we  make more conservative\footnote{Throughout the text, we use ``conservative'' to imply biasing the GRB distribution towards lower redshifts.  This is conservative since it mimics the effects of WDM, thereby resulting in weaker constraints on the particle mass.} assumptions throughout the analysis, such as normalizing  the SFR-GRB ratio at high redshifts (thereby using a shorter, more accurate lever arm which minimizes modeling uncertainty),  using  an unbiased luminosity function and allowing the  SFR-GRB ratio to  evolve with redshift. Finally, we study the effectiveness of  future observations in improving the current constraints.

Current limits on dark matter masses, $m_{\rm x}$, are motivated by several observations. The Lyman-$\alpha$ forest implies $m_{\rm x} \gtrsim 1$ keV \citep[e.g.,][]{Viel2008} and  $m_{\nu s} > $  8 keV  for sterile neutrinos \citep{Seljak2006,Boyarsky2009b}. Likewise,   WDM models with a too warm candidate ($m_{\rm x} < 0.75$ keV) cannot simultaneously reproduce the stellar mass function and the Tully-Fisher relation \citep{Kang2013}. Also, the fact that reionization occurred at $z\gtrsim6$ implies $m_{\rm x} \gtrsim 0.75$ keV \citep{Barkana2001}.   
However, all of these limits are strongly affected by a degeneracy between astrophysical (i.e. baryonic) processes and the dark matter mass. Our approach in this work is more robust, driven only by the shape of the redshift evolution of the $z>4$ SFR.  Furthermore,  it is important to note that the SFR is exponentially attenuated at high-redshifts in WDM models.  Since astrophysical uncertainties are unable to mimic such a rapid suppression, probes at high-redshifts (such as GRBs and reionization) are powerful in constraining WDM cosmologies.

 The outline of this paper is as follows. In \S 2 we discuss how we derive the dark matter halo mass function and SFR in WDM and CDM models.  In \S 3 we derive the corresponding GRB redshift distribution. In \S 4  we discuss the  adopted observed GRB sample. In \S  5 we present our analysis and main results. In \S 6, we discuss possible future constraints using a theoretical mock sample. Finally,  in \S 7,  we present our conclusions.\footnote{Throughout the paper, we adopt the cosmological parameters from  WMAP9  \citep{Hinshaw2012},  $\Omega_m = 0.264, \Omega_{\Lambda} = 0.736$, $n_s$ = 0.97, $\sigma_8=0.8$ and $H_0 = 71$ km s$^{-1}$ Mpc $^{-1}$.}


\section{Structure formation in a WDM dominated  universe}

Massive neutrinos from the standard model (SM) of particle physics were one of the first dark matter candidates. However, structures formed in this paradigm are incompatible with observations. Other alternative dark matter candidates usually imply   an extension of the SM.  The DM  particle candidates  span  several order of magnitude in  mass \citep{Boyarsky2009a}:  axions with a mass of $\sim 10^{-6}$ eV, first  introduced  to solve the problem of CP violation in particle physics, supersymmetric (SUSY) particles (gravitinos, neutralinos, axinos) with  mass in the range $\sim$ eV-GeV, superheavy dark matter, also called \textit{Wimpzillas}, (also considered as a possible   solution to the problem of cosmic rays observed above the GZK cutoff), Q-balls, and sterile neutrinos with mass $\sim$  keV range, just to cite a few. For a  review about dark matter candidates see \cite{Bertone2005}. 
Two  promising  candidates for warm dark matter   are the sterile neutrino \citep{Dodelson1994,Shaposhnikov2006} and gravitino \citep{Ellis1984,Moroi1993,Kawasaki1997,Primack2003,Gorbunov2008}.

In WDM models the  growth of density perturbations  is  suppressed on scales smaller than the free streaming length.  The lighter the WDM particle,  the larger the scale below which  the power spectrum is suppressed.  In addition to this power-spectrum cutoff, one must also consider the residual particle velocities.  As described in \citet{Barkana2001}, these act as an effective pressure, slowing the early growth of perturbations.  Bellow we describe how we include these two effects in our analysis (for more information, please see \citealt{Barkana2001, Mesinger2005}).

\subsection{Power spectrum cutoff}

 The free-streaming scale, below which  the linear perturbation amplitude is suppressed \citep[e.g.,][]{Colombi1996,Bode2001,Viel2005},  is given by  the comoving scale 

\begin{equation}
R_{\rm fs} \approx 0.11\left(\frac{\Omega_{\rm x}h^2}{0.15}\right)^{1/3}\left(\frac{m_{\rm x}}{\rm keV} \right)^{-4/3} \rm Mpc,
\end{equation}
where $\Omega_{\rm x}$ is the total energy density contained
in WDM particles relative to the critical density, $h$ is the Hubble constant in units of 100 $\rm km s^{-1} Mpc^{-1}$, and  $m_{\rm x}$ is the WDM particle mass.

The resulting modification  of the matter power spectrum can be computed by multiplying the CDM power spectrum $P_{\rm CDM}(k)$ by an additional transfer function \citep{Bode2001}:

\begin{equation}
P_{\rm WDM}(k) = P_{\rm CDM}(k)\left[1+(\epsilon k)^{2\mu}\right]^{-5\mu},
\label{eq:Pk}
\end{equation}
where $\mu = 1.12$ and 
\begin{equation}
\epsilon = 0.049 \left(\frac{\Omega_{\rm x}}{0.25}\right)^{0.11}\left(\frac{m_{\rm x}}{\rm keV}\right)^{-1.11}\left(\frac{h}{0.7}\right)^{1.22}h^{-1} \rm Mpc.
\end{equation}

\subsection{Effective pressure}

Structure formation in WDM models will be  further suppressed by the residual velocity dispersion of the WDM particles, which delays  the growth of perturbations.  As shown in \citet{Barkana2001} and \citet{Mesinger2005}, this ``effective pressure'' is of comparable importance to the power spectrum cutoff in determining the WDM mass functions. The pressure   can be incorporated  in the halo  mass function   by raising the critical linear extrapolated overdensity threshold at collapse,  $\delta_{c}(M,z)$.  Using spherically symmetric hydrodynamics simulations,  and exploiting the analogy between the WDM effective pressure and gas pressure, \citet{Barkana2001} computed the modified $\delta_{c}(M, z)$.  They showed that one can define an 
effective WDM Jeans mass, i.e., the mass scale below which collapse is significantly delayed by the pressure.  We will denote this scale as:
\begin{equation}
M_{\rm WDM} \approx 1.8\times10^{10}\left(\frac{\Omega_{\rm x}h^2}{0.15}\right)^{1/2}\left(\frac{m_{\rm x}}{1 \rm keV}\right)^{-4} ~ , 
\label{eq:MWDM}
\end{equation}
where we make the standard assumption of a fermionic spin 1/2 particle.  Note that equation (\ref{eq:MWDM}) differs from the original one proposed by \citet{Barkana2001} by a factor of 60.  With this adjustment, we find that the collapsed fractions, $F_{\rm {coll}}(z)$,  simulated assuming a sharp mass cutoff at $M_{\rm WDM}$ are in good agreement with the full random-walk procedure (assuming a more gradual rise in $\delta_c(M)$) of \citet{Barkana2001} (see Fig. \ref{fig:colfrac}).  The former approach has the advantage of being considerably faster and simpler.

\subsection{Collapsed fraction of haloes and cosmic star formation}

We use the Sheth-Tormen mass function, $f_{\rm ST}$, \citep{sheth1999} to estimate  the number density of dark matter haloes, $n_{\rm ST}(M,z)$, with mass greater than  $M$:
\begin{equation}
f_{\rm ST} = A\sqrt{\frac{2a_1}{\pi}}
\left[1+\left(\frac{\sigma^2}{a_1\delta_{c}^2}\right)^p\right]
\frac{\delta_{c}}{\sigma}\exp{\left[-\frac{a_1\delta_{c}^2}{2\sigma^2}\right]}, 
\end{equation}
where $A=0.3222, a_1 = 0.707, p = 0.3$ and  $\delta_c = 1.686$. The mass function $f_{\rm ST}$ can be related to $n_{\rm ST}(M,z)$ by 
\begin{equation}
f_{\rm ST} = \frac{M}{\rho_{\rm m}}\frac{dn_{\rm ST}(M,z)}{d\ln{\sigma^{-1}}}, 
\end{equation}
 where $\rho_{\rm {m}}$ is the total mass density of the background Universe. 
The variance of the linearly extrapolated density field $\sigma (M,z)$ is given by 
\begin{equation}
\sigma^{2}(M,z) = \frac{b^2(z)}{2\pi^2}\int_{0}^{\infty} k^2P(k)W^2(k,M)dk,
\end{equation}
where $b(z)$ is the growth factor of linear perturbations normalized 
to $b = 1$ at the present epoch, and  $W(k,M)$ is a real space top-hat filter. In order to calculate the CDM power spectrum $P_{\rm CDM}(k)$, we use the \textsc{CAMB} code\footnote{http://camb.info/}. For WDM models, we compute the power spectrum, $P_{\rm WDM} (k)$, with equation (\ref{eq:Pk}).

\begin{figure}
\psfrag{1e6}[c][c]{$10^{6}$}
\psfrag{1e7}[c][c]{$10^{7}$}
\psfrag{1e8}[c][c]{$10^{8}$}
\psfrag{1e9}[c][c]{$10^{9}$}
\psfrag{1e10}[c][c]{$10^{10}$}
\psfrag{1e11}[c][c]{$~10^{11}$}
     \psfrag{1e12}[c][c]{$~10^{12}$}
       \psfrag{1e13}[c][c]{$10^{13}$}
        \psfrag{1e14}[c][c]{$10^{14}$}
     \psfrag{1e15}[c][c]{$10^{15}$}
     \psfrag{1e16}[c][c]{$10^{16}$}
\psfrag{0}[c][c]{$0$}
\psfrag{1}[c][c]{$1$}
     \psfrag{2}[c][c]{$2$}
     \psfrag{3}[c][c]{$3$}
     \psfrag{4}[c][c]{$4$}
     \psfrag{5}[c][c]{$5$}
     \psfrag{6}[c][c]{$6$}
     \psfrag{7}[c][c]{$7$}
      \psfrag{8}[c][c]{$8$}
           \psfrag{9}[c][c]{$9$}
      \psfrag{10}[c][c]{$10$}
      \psfrag{11}[c][c]{$11$}
      \psfrag{12}[c][c]{$12$}
      \psfrag{13}[c][c]{$13$}
      \psfrag{14}[c][c]{$14$}
      \psfrag{15}[c][c]{$15$}
     \psfrag{z}[c][c]{$z$}
     \psfrag{ylabel}[c][c]{$\rm M_{h}(M_{\odot})~~~~$}
\includegraphics[trim = 0mm 4mm 4mm 18mm, clip, width=1\columnwidth]{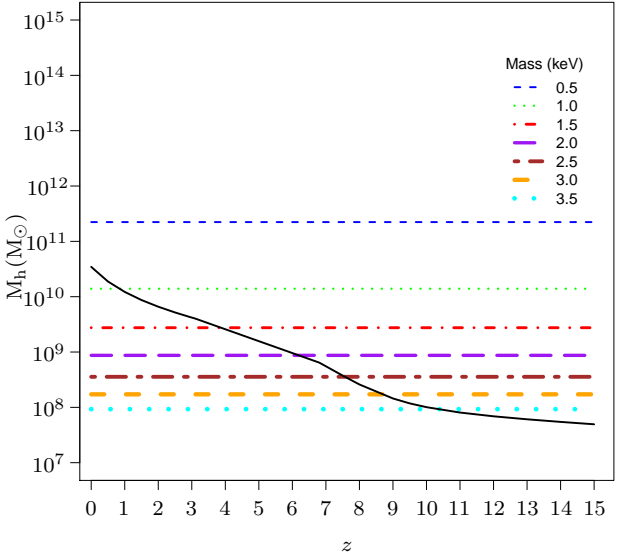}
\caption{The minimum halo masses capable of hosting star-forming galaxies.  The solid black line corresponds to the astrophysical limit, $M_{\rm gal}$,  from \citet{SM2013b}.  
The horizontal lines correspond to the cosmological cutoffs in WDM models, $M_{\rm WDM}$ ($m_{\rm x}$ = 0.5,1,1.5,2,2.5,3,3.5  keV from top to bottom, respectively).  }
\label{fig:massrange}
\end{figure}

The fraction of mass inside collapsed halos, $F_{\rm {coll}}(>M_{\rm min}, z)$, is then given by:
\begin{equation}
F_{\rm {coll}}(z)=\frac{1}{\rho_{\rm {m}}}\int_{M_{\rm min}}^{\infty} dMMn_{\rm {ST}}(M,z), 
\label{fcol}
\end{equation}
and the minimum mass is estimated  as
  \begin{equation}
 M_{\rm min} = \rm Max[M_{\rm gal}(z), M_{WDM}(m_{\rm x})],
 \end{equation}
 where  $M_{\rm gal}$ corresponds to the minimum halo mass capable of hosting star forming galaxies. We use $M_{\rm  gal}  \equiv \bar{M}_{min}$ from equation (11) in \citet{SM2013b}, who present a physically-motivated expression for the evolution of $M_{\rm gal}$  which takes  into account  a gas cooling criterion  as well as radiative feedback from an ionizing UV background (UVB) during inhomogeneous reionization\footnote{Using $M_{\rm gal}(z)$ from \citet{SM2013b} for WDM models is not entirely self-consistent, since UVB feedback effects should be smaller in WDM models.  However, this effect is smaller than other astrophysical uncertainties.  Most importantly, our conclusions are driven by models which at high redshift have $M_{\rm WDM} > M_{\rm gal}$, and are therefore insensitive to the actual choice of $M_{\rm gal}$ (see Fig. 2).}.
  In other words, $M_{\rm gal}$ is set by astrophysics, whereas $M_{\rm WDM}(m_{\rm x})$ is set by the particle properties of  dark matter.  In Fig. \ref{fig:massrange}, we show these two limits as a function of redshift.  
   As we shall see, our conclusions are derived from the regime where $M_{\rm WDM} > M_{\rm gal}$,  and hence they are  not sensitive to the exact value of $M_{\rm gal}$. Another point worth highlighting  is that models with $m_{\rm x} \gtrsim 3.5$ keV have $M_{\rm WDM}(m_{\rm x}) \geq M_{\rm gal}$  up to $z \approx 11$. Thus, any constraint beyond $\sim 3.5$keV will no longer be robust, since galaxy formation can be suppressed  even in CDM, bellow halo masses comparable to  $M_{\rm WDM}$.       Therefore, observations  at much higher redshifts are required to obtain  tighter constraints. 

\begin{figure}
\psfrag{1e-08}[c][c]{$-8$}
\psfrag{1e-06}[c][c]{$-6$}
         \psfrag{1e-04}[c][c]{$-4$}
          \psfrag{1e-02}[c][c]{$-2$}
                    \psfrag{1e+00}[c][c]{$0$}
    \psfrag{0}[c][c]{$0$}
     \psfrag{2}[c][c]{$2$}
     \psfrag{4}[c][c]{$4$}
     \psfrag{6}[c][c]{$6$}
      \psfrag{8}[c][c]{$8$}
            \psfrag{9}[c][c]{$9$}
      \psfrag{10}[c][c]{$10$}
     \psfrag{z}[c][c]{$z$}
     \psfrag{Fcol}[c][c]{$\log{\rm F_{\rm coll}(>M, z)}~~~$}
\includegraphics[width=1\columnwidth]{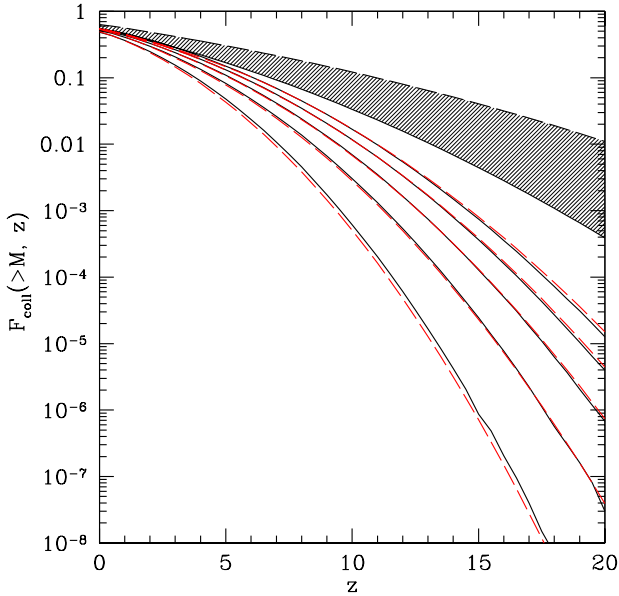}
\caption{Fraction of the total mass collapsed into haloes of mass $>M_{\rm min}$ as a function of redshift, $F_{\rm coll}(>M_{\rm min}, z)$.  The shaded region shows the collapsed fraction in CDM, with a range of low-mass cutoffs corresponding to virial temperatures 300 K $< T_{\rm vir} < 10^4$ K. The other curves correspond to WDM particle masses of  $m_{\rm x}$  =   3.0, 2.5, 2.0, 1.5,  and 1.0 keV (top to bottom). The figure is an adapted version of Fig. 2 from \citet{Mesinger2005} (computed using their cosmology), serving to motivate equation (\ref{eq:MWDM}).  Values of $F_{\rm coll}(>M_{\rm min},z)$ computed according to the full random walk procedure used in \citet{Mesinger2005} are shown as solid black curves, while the approximation of a sharp cutoff at $M_{\rm WDM}$ used in this work corresponds to the red dashed curves.
\label{fig:colfrac}}
\end{figure}

In figure \ref{fig:colfrac},  we plot the fraction of the total mass collapsed into haloes of mass $>M_{\rm min}$, $F_{\rm coll}(>M_{\rm min},z)$.  The shaded region shows the collapsed fraction in CDM, with a range of low-mass cutoffs corresponding to virial temperatures 300 K $<T_{\rm vir}<10^4$ K. The other curves correspond to WDM particle masses of $m_{\rm x}$ = 3.0, 2.5, 2.0, 1.5, and 1.0 keV (top to bottom).  This figure is analogous to Fig. 2 in \citet{Mesinger2005}, serving to motivate equation (\ref{eq:MWDM}).  The fractions $F_{\rm coll}(>M_{\rm min},z)$,  computed according to the full random walk procedure used in \citet{Mesinger2005},  are shown as solid black curves, while the approximation of a sharp cutoff at $M_{\rm WDM}$ (eq.\ref{eq:MWDM}) corresponds to the red dashed curves.

The comoving star formation rate  as a function of redshift is assumed to be  proportional  to $dF_{\rm {coll}}/dt$:
\begin{equation}
SFR(z)  \propto \frac{\rm {d}F_{\rm {coll}}}{dt}.
\label{SFRH}
\end{equation}
The proportionality constant is irrelevant for our analysis, as it is subsumed in our normalization procedure below.


\section{Theoretical redshift  distribution of GRBs}
\label{sec:CDFG}

\label{sec_GRB_rate}
Under the hypothesis that the formation rate of long GRBs (LGRBs; duration longer than 2 sec) follows the SFR (e.g., \citealt{totani1997,campisi2010, bromm2006, rafael2011}),  the comoving rate of GRBs,   $\Psi_{\rm GRB}$,  can be expressed as 
\begin{equation}
\Psi_{\rm GRB}(z) =\zeta_{0}(1+z)^{\alpha} SFR(z).
\label{eq:psigrb}
\end{equation}
Here $\zeta_{0}$ is a constant that includes the absolute conversion from the SFR to the GRB rate.
 The evolutionary trend described by $\alpha$ may arise from several mechanisms \citep[e.g.,][]{kistler2009}, with a possible explanation provided by the GRB preference for low-metallicity environments\footnote{Since host galaxies of long duration GRBs are often observed to be metal poor, 
 several studies have tried to connect the origin of long GRBs with the metallicity of their progenitors \citep[e.g.,][]{meszaros2006,woosley2006,Salvaterra2007,Salvaterra2009,campisi2011b}. Such a connection is physically-motivated since core-collapse models could
not generate a long-GRB without the progenitor system having low metallicity
 \citep[e.g.,][]{Hirschi2005,Yoon2005,woosley2006}.  On the other hand, several authors report observations of GRBs in high metallicity  environments \citep[e.g.,][]{Levesque2010,Kruhler2012}, suggesting that GRB hosts are not necessarily metal poor.
Despite the apparent preference of GRBs towards metal-poor hosts, there is no clear cutoff in metallicity,  above which GRB formation should be suppressed.}.
  A metallicity threshold seems to provide a natural explanation for the observed value of $\alpha \approx 0.5-1$ at low redshifts ($z \leqslant 4$; e.g. \citealt{Robertson2012}).  Such a metallicity threshold increases with redshift, whereas the characteristic halo mass decreases with redshift.  In such a scenario, a value of $\alpha = 0$ would be appropriate for our high-redshift ($z>4$) analysis.  Nevertheless, we conservatively keep $\alpha$ as a free parameter.

The observed number of GRBs in the range $z_1 \leqslant z \leqslant z_2$,  $N(z_1,z_2)$, can be expressed by 
\begin{equation}
N(z_1,z_2) = K\int_{z_1}^{z_2} \frac{dN_{\rm GRB}}{dz'}dz',
\label{eq:cum_numb}
\end{equation}
with 
\begin{equation}
\frac{dN_{\rm GRB}}{dz} = \Psi_{\rm GRB}(z)\frac{\Delta t}{1+z}\frac{dV}{dz}\mathcal{I}(z), 
\label{eq:DNDZ}
\end{equation}
where the parameter $K$   accounts for the efficiency of  finding  GRBs and measuring   their redshift   (e.g., area coverage, the survey flux limit, beaming factor of GRBs, etc) \footnote{There are several selection effects known to mask the true GRB redshift distribution, e.g.:  (i) the host galaxy dust extinction; (ii) the redshift desert (a redshift span, $1.4 < z < 2.5$, in which it is difficult to measure absorption and emission spectra); (iii) Malmquist bias; and (iv) the difference between redshift measurements techniques \citep[e.g.,][]{Coward2012}.  We therefore expect $K$ to be redshift dependent, and this evolution is subsumed in our parameter $\alpha$ above. Most of the above effects (e.g. obtaining GRB redshifts, dust extinction,  Malmquist bias) result in biasing the observed sample towards low redshifts.  Since we calibrate the proportionality between the GRB rates and SFRs at $z\sim 3-4$, we likely overestimate the efficiency in the redshift determination of $z>4$ GRBs.  Therefore, we expect that even a non-evolving $K$ (i.e. our results for $\alpha=0$) would be a conservative assumption.},
  $dV/dz$ is the comoving  volume element per unit redshift,  $\Delta t$ is the time interval  in the observer rest frame\footnote{Which in our case, corresponds to 5 yrs of observation time by \textit{Swift}. Again, the value is not relevant for our purposes,  since it is subsumed in the normalization.  },  and  $\mathcal{I}(z)$ is the integral over   the GRB luminosity function (LF),  $p(L)$, 
\begin{equation}
\mathcal{I}(z)=\int_{\log{L_{lim}(z)} }^{\infty}p(L)d\log{L}.
\end{equation}
To remove the dependence on proportionality constants, we  construct  cumulative distribution functions (CDFs) of GRBs  over the redshift range $z_i < z < z_{\rm max}$:
\begin{equation} 
N (< z|z_{\rm max}) = \frac{N(z_i,z)}{N(z_i,z_{\rm max})}.
\end{equation}

 The expected number of observed GRBs in a given redshift interval, for each combination $\pmb{\theta}\equiv \{m_{\rm x},\alpha\}$,  can be written as 
\begin{equation}
N(z_1,z_2;\pmb{\theta}) = \zeta_{0;\pmb{\theta}} \int_{z_1}^{z_2}(1+z)^{\alpha} SFR(z;m_{\rm x})\frac{\Delta t}{1+z}\frac{dV}{dz}\mathcal{I}(z), 
\label{eq:Nz1z2}
\end{equation}
where $\zeta_{0;\pmb{\theta}}$ is {\it normalized so that  each model recovers  the observed GRB rate, $N_{\rm obs}(z_1,z_2)$,  at $z\approx 4$}, 
\begin{equation}
\zeta_{0;\pmb{\theta}} = \frac{N_{\rm obs}(3,4)}{\int_{3}^{4}(1+z)^{\alpha} SFR(z;m_{\rm x})\frac{\Delta t}{1+z}\frac{dV}{dz}\mathcal{I}(z)}.
\label{eq:zeta0}
\end{equation}
 $N_{\rm obs}(3,4)$ is equal to 24 (7) for S1 (S2) samples respectively. Note that normalizing at a relatively high-$z$ ($z=$3--4) is indeed a   conservative choice.   If we had normalized at lower redshifts, say $z \sim1-2$, the absolute number of GRBs predicted by $m_{\rm x} = 0.5$  keV and CDM,  between $z \sim  3-4$, would already diverge by a factor of  $\sim 2$ for $\alpha$ values in the range $0-2$.  
In addition to being conservative, normalizing at high-redshifts allows us to have a relatively short lever arm over which our simple scaling, $SFR\propto dF_{\rm coll}/dt$, is presumed to be accurate.  At lower redshifts, mergers and AGN feedback (missing from our model) are expected to be important in determining the SFR.  Hence to normalize our CDFs, we chose the largest redshift at which our sample is reasonably large (see below).  It is important to note however that our main results are based on the $z>4$ CDFs, which unlike the predictions for the absolute numbers of bursts, {\it do not depend on our choice of normalization.}



\section{GRB sample}

Our  LGRB data is taken from \citet{Robertson2012}, corresponding to a compilation from  the samples presented in \citet{Butler2007,Perley2009,Butler2010,Sakamoto2011,Greiner2011}, and \citet{Kruhler2011}. It  includes only GRBs  before the end of the Second Swift BAT GRB Catalog, and is comprised of 152 long GRBs  with redshift measurements.
 It is important to consider the completeness of the sample. Several efforts have been made to construct  a redshift-complete GRB sample  \citep[e.g.,][]{Greiner2011,Salvaterra2012}. However, to do so, many GRBs with measured redshifts are excluded. Such requirements are  even more severe for  high-$z$ bursts, which makes them of little use for our purposes. 
To explore the dependence of our results on a possible bias in the GRB redshift distribution,  we construct two samples: (S1) we use  a luminosity function based on the $z<4$ subsample (consisting of 136 GRBs);
 and (S2) we use a subsample with isotropic-equivalent luminosities bright enough to be observable  up to high redshifts (comprised of 38 bursts).  The two  samples are summarized in table \ref{tab:cases}.  Since there is a degeneracy between a biased SFR-GRB relation and a redshift-dependent LF, we implicitly  assume that any unknown  bias will be subsumed  in the value of the $\alpha$ parameter. 

\subsection{Luminosity function sample (S1)}

\begin{figure}
\psfrag{y}[c][c]{Frequency}
\psfrag{x}[c][c]{$ \log_{10}{(\rm L_{iso}/\rm erg ~s^{-1})}$}
\psfrag{46}[c][c]{$46$}
\psfrag{48}[c][c]{$48$}
\psfrag{50}[c][c]{$50$}
\psfrag{52}[c][c]{$52$}
\psfrag{54}[c][c]{$54$}
\psfrag{56}[c][c]{$56$}
\psfrag{0.0}[c][c]{$0$}
\psfrag{0.1}[c][c]{$0.1$}
\psfrag{0.2}[c][c]{$0.2$}
\psfrag{0.3}[c][c]{$0.3$}
\psfrag{0.4}[c][c]{$0.4$}
\psfrag{0.5}[c][c]{$0.5$}

\includegraphics[trim = 0mm 4mm 4mm 18mm, clip,width=1\columnwidth]{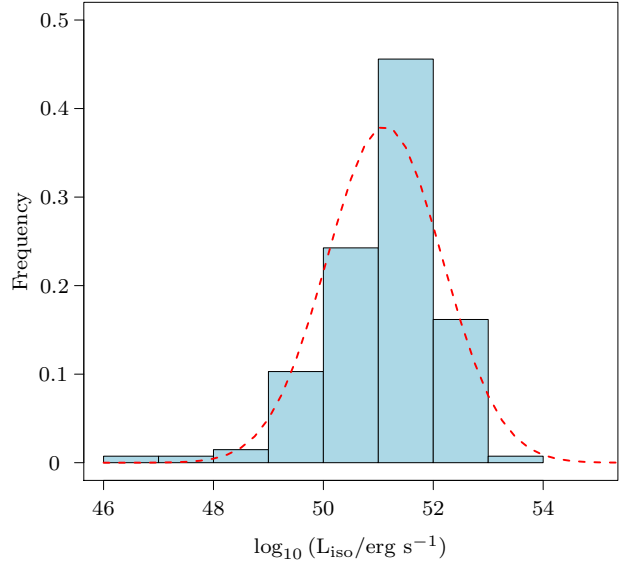}
\caption{Frequency (i.e. fraction in bin) of GRB luminosities for the $z<4$ subsample used to construct the LF used for S1 (see text for details). The red dashed line represents the  best-fit LF.}
\label{fig:SWIFT_hist}
\end{figure}

The number of GRBs detectable by any given instrument depends on the 
specific flux sensitivity threshold and  the intrinsic 
isotropic LF of the  GRBs.  In figure \ref{fig:SWIFT_hist}, we show the distribution of  log-luminosities for $z<4$ GRBs, which can be well described by a normal distribution
\begin{equation}
\label{LF}
p(L) = p_*e^{-(L-L^*)^2}/2\sigma^{2}_L. 
\end{equation}
The values  $L = \log{\rm L_{iso}/erg~s^{-1}}$;  $L^* = \log{10^{51.16}}$,  $\sigma_L = 1.06$ and $p_* = 1.26$ are  estimated by   maximum likelihood optimization.  The luminosity threshold is given by
\begin{equation}
\label{eq:limit}
  L_{\rm lim} = 4\pi\, d_{\rm L}^{2}\, F_{\rm lim}, 
\end{equation}
where $d_L$ is the luminosity distance. Consistent with previous works (e.g. \citealt{li2008}), we set a bolometric energy flux limit $F_{\rm lim} = 1.15 \times 10^{-8} {\rm erg}~ {\rm cm}^{-2}~ {\rm s}^{-1}$  for \textit{Swift} by using the smallest luminosity of the sample.
Due to Malmquist bias, our  fitted LF is likely biased towards  high luminosities. To minimize this problem and increase the sample completeness, we fit our LF using only $z\leqslant 4$ GRBs,  which we show in Fig. \ref{fig:SWIFT_hist}.  We reiterate that the Malmquist bias only serves to make our results even more conservative by predicting a flatter redshift distribution of GRBs.

\subsection{Luminosity-limited sample (S2)}

Another approach,  less dependent on the  LF parametrization and the Malmquist bias,  is to construct  a luminosity-limited subsample of the observed bursts bright enough to be seen at the highest redshift of interest. Assuming  that the LF does not evolve with redshift,  this subset  would be proportional to the total number of bursts at any given redshift.

\begin{figure}
\psfrag{y}[c][c]{$\rm L_{iso} (erg~s^{-1})$}
\psfrag{z}[c][c]{$z$}
\psfrag{1e+53}[c][c]{$~~10^{53}$}
\psfrag{1e+51}[c][c]{$~~~10^{51}$}
\psfrag{1e+49}[c][c]{$~~10^{49}$}
\psfrag{1e+47}[c][c]{$~~10^{47}$}
\psfrag{0}[c][c]{$0$}
\psfrag{2}[c][c]{$2$}
\psfrag{4}[c][c]{$4$}
\psfrag{6}[c][c]{$6$}
\psfrag{8}[c][c]{$8$}
\psfrag{10}[c][c]{$10$}
\includegraphics[trim = 0mm 4mm 4mm 18mm, clip,width=1\columnwidth]{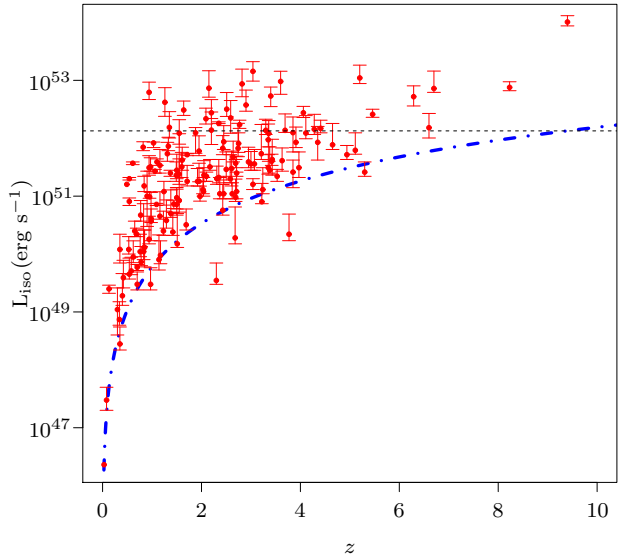}
\caption{Isotropic luminosities,  $L_{\rm iso}$,  of 152 {\it Swift} gamma-ray bursts as a function of  $z$  from the compilation  of \citealt{Robertson2012}. The blue dot-dashed line  approximates the effective \textit{Swift} detection threshold (eq. \ref{eq:limit}). The black dashed horizontal line represents the luminosity limit of  $L_{\rm iso}>1.34\times10^{52}$ $\rm ergs~s^{-1}$, used to define our S2 subsample.   }
\label{fig:SWIFT_LF}
\end{figure}

  In figure \ref{fig:SWIFT_LF}, we show the redshifts and isotropic luminosities of our entire sample.  The dot-dashed blue line corresponds to the  effective \textit{Swift} detection threshold.
For our luminosity-limited sample, we only use GRBs with  isotropic-equivalent luminosities $L_{\rm iso} \geqslant 1.34\times10^{52} \rm ergs~ s^{-1}$, which comprise all GRBs observable up to $z\sim 9.4$.   Hereafter, all calculations will correspond to either the complete (LF derived) sample (S1),  or the luminosity-limited sample (S2).

\section{Observational Constraints}
\label{sec:obs}

In this section, we test the WDM models by comparing  the  predicted absolute detection rates  of bursts as well as the   CDFs  with the observed samples.
We consider   3 different ranges of $\alpha$:  (i) a constant SFR-GRB relation,  $\alpha = 0 $ (case 0, C0); (ii) $-1<  \alpha < 2$ (case 1, C1); and (iii) a flat prior over  $-\infty < \alpha < \infty$ (case 2, C2)\footnote{More precisely, C2 was run over the interval $-3 < \alpha < 12$, which was more than sufficient to capture the likelihood decreasing to $\rightarrow0$ in the tails of the distribution (see Fig. \ref{fig:alpha_contour1}).}.  All cases are summarized in table \ref{tab:cases}.

 \begin{table}
\caption{Set of cases considered in our analysis.}
\begin{center}
\begin{tabular}{lccc}
\hline
$L_{\rm lim}(\rm ergs~ s^{-1}$) & $\alpha = 0$ & $-1<  \alpha < 2$ & $-\infty<\alpha <\infty$\\
\hline
\hline
$L_{\rm lim} \geqslant 0$ &   S1C0          &S1C1  & S1C2\\
 $L_{\rm lim} \geqslant 1.34\times10^{52} $   &S2C0& S2C1 & S2C2\\
\label{tab:cases}
\end{tabular}
\end{center}
\label{default}
\end{table}%

\subsection{Absolute  detection rate of bursts}
\label{sec:abs_constraints}

In tables \ref{tab:summary}-\ref{tab:summary2}, we present the absolute number of GRBs  at high redshifts in CDM and WDM models with particle masses of 0.5-3.5 keV, as well as the actual number in our sample observed with {\it Swift}.  All models are normalized to yield the observed number of bursts at $3<z<4$, as described in equation (\ref{eq:zeta0}) and the associated discussion.

 As expected, models with small WDM particle masses predict a rapidly decreasing GRB rate towards high redshifts.  This exponential suppression can in some cases be partially compensated by an increasing GRB-to-SFR rate  (i.e. $\alpha > 0$). For the S1C1 case, models with $m_{\rm x} \geqslant 2.5$  keV  show a good agreement with  \textit{Swift }observations for  $ 0 < \alpha < 1$, though values of $\alpha \sim 2$ are a better fit to the observations at $z > 8$.  For the case S2C1, all models with $m_{\rm x} \geqslant 2.5$  keV seem to be consistent with data for $\alpha \sim 1-2$. In both cases, the  two observed bursts  in the interval $8 < z < 10$, are already  at odds with  1.5 keV $< m_{\rm x} <$ 2.5 keV models.  Finally, we see that models with  $m_{\rm x} \leqslant 1$ keV predict a dearth of GRBs at $z>6$, which is inconsistent with current observations.  Extreme models with $m_{\rm x} \sim 0.5$ keV already fail at intermediate redshifts ($4 < z < 6$), even for values of $\alpha$ as high as two.

\begin{table*}
\caption{Absolute number of GRBs per redshift interval  predicted by each model for S1C1 sample.   }
\label{tab:summary}
\begin{center}
\begin{tabular}{lccc|ccc|cccc|}
\hline
Model  & \multicolumn{3}{c}{N(4,6)} & \multicolumn{3}{c}{N(6,8)} &\multicolumn{3}{c} {N(8,10)}   \\
\hline
\hline

& $\alpha = 0$& $\alpha=1$ & $\alpha = 2$ &$\alpha=0$& $\alpha=1$ & $\alpha = 2$ & $\alpha=0$& $\alpha=1$ & $\alpha = 2$ \\
\hline
 $m_{\rm x}$ = 0.5 keV  &3.18 & 3.94 & 4.88 & 0.01 & 0.02 & 0.04 & $1.0\times10^{-5}$& $2.2\times10^{-5}$ & $4.6\times10^{-5}$\\
 $m_{\rm x}$ = 1.0 keV  &9.34 & 11.82 & 15.00 & 0.42 & 0.71 & 1.21 & 0.01 & 0.02 & 0.04 \\ 
$m_{\rm x}$ = 1.5 keV & 14.84 & 19.10 & 24.67 & 1.51 & 2.58 & 4.42 & 0.08 & 0.17 & 0.36 \\
$m_{\rm x}$ = 2.0 keV & 14.31 & 18.36 & 23.65 & 2.31 & 4.01 & 6.96 & 0.20 & 0.43 & 0.93 \\ 
$m_{\rm x}$ = 2.5 keV &  14.44 & 18.54 & 23.90 & 2.09 & 3.65 & 6.38 & 0.39 & 0.83 & 1.80 \\ 
$m_{\rm x}$ = 3.0 keV & 14.51 & 18.62 & 24.01 & 2.04 & 3.54 & 6.16 & 0.46 & 1.00 & 2.20 \\
$m_{\rm x}$ = 3.5 keV& 14.56 & 18.69 & 24.09 & 2.04 & 3.55 & 6.18 & 0.44 & 0.97 & 2.13 \\ 
CDM   & 14.56 & 18.69 & 24.10 & 2.04 & 3.55 & 6.18 & 0.45 & 1.00 & 2.19 \\
\textit{Swift}                    &    11        &     11                                            &    11  &3        & 3                                           & 3  & 2                                    &                       2                          &       2 \\
\hline
\end{tabular}\\
\end{center}
\end{table*}

\begin{table*}
\caption{Absolute number of GRBs per redshift bin  predicted by each model for S2C1 sample.  }
\label{tab:summary2}
\begin{center}
\begin{tabular}{lccc|ccc|cccc|}
\hline
Model  & \multicolumn{3}{c}{N(4,6)} & \multicolumn{3}{c}{N(6,8)} &\multicolumn{3}{c} {N(8,10)}   \\
\hline
\hline

& $\alpha = 0$& $\alpha=1$ & $\alpha = 2$ &$\alpha=0$& $\alpha=1$ & $\alpha = 2$ & $\alpha=0$& $\alpha=1$ & $\alpha = 2$ \\
\hline
 $m_{\rm x}$ = 0.5 keV  &1.33 & 1.65 & 2.05 & 0.01 & 0.02 & 0.03 & $1.5\times10^{-5}$& $3.1\times10^{-5}$ & $6.6\times10^{-5}$\\
 $m_{\rm x}$ = 1.0 keV  &4.10 & 5.23 & 6.70 & 0.34 & 0.59 & 1.00 & 0.01 & 0.03 & 0.06 \\ 
$m_{\rm x}$ = 1.5 keV &    6.75 & 8.78 & 11.46 & 1.27 & 2.19 & 3.76 & 0.12 & 0.26 & 0.55 \\ 
$m_{\rm x}$ = 2.0 keV &  6.47 & 8.40 & 10.94 & 2.02 & 3.52 & 6.14 & 0.31 & 0.67 & 1.44 \\  
$m_{\rm x}$ = 2.5 keV & 6.54 & 8.49 & 11.07 & 1.86 & 3.27 & 5.77 & 0.60 & 1.30 & 2.81 \\
$m_{\rm x}$ = 3.0 keV&   6.57 & 8.53 & 11.12 & 1.79 & 3.13 & 5.49 & 0.74 & 1.63 & 3.58 \\
$m_{\rm x}$ = 3.5 keV&   6.59 & 8.56 & 11.16 & 1.80 & 3.14 & 5.51 & 0.72 & 1.58 & 3.48 \\
CDM   &   6.60 & 8.56 & 11.16 & 1.80 & 3.15 & 5.51 & 0.74 & 1.63 & 3.58 \\
\textit{Swift } &6 & 6 & 6 &3 & 3 & 3&  2&  2& 2 \\
\hline
\end{tabular}\\
\end{center}
\end{table*}

\subsection{The redshift  distribution  of $z>4$ bursts }
\label{sec:CDF}

Although the absolute rate of bursts is the simplest prediction, it is dependent on the normalization factor between the SFR and GRB rate at $3<z<4$.  Hence, for the remainder of the paper, we focus on comparing the theoretical and observed $z>4$ CDFs.  The CDFs are not dependent on normalization factors and are therefore more conservative and robust predictions.

\begin{figure}
\psfrag{Fn(x)}[c][c]{CDF}
\psfrag{x}[c][c]{$z$}
\psfrag{4}[c][c]{$4$}
\psfrag{5}[c][c]{$5$}
\psfrag{6}[c][c]{$6$}
\psfrag{7}[c][c]{$7$}
\psfrag{8}[c][c]{$8$}
\psfrag{9}[c][c]{$9$}
\psfrag{10}[c][c]{$10$}
\psfrag{0.0}[c][c]{$0$}
\psfrag{0.2}[c][c]{$0.2$}
\psfrag{0.4}[c][c]{$0.4$}
\psfrag{0.5}[c][c]{$0.5$}
\psfrag{0.6}[c][c]{$0.6$}
\psfrag{0.8}[c][c]{$0.8$}
\psfrag{1.0}[c][c]{$1.0$}
\psfrag{1.5}[c][c]{$1.5$}
\psfrag{2.0}[c][c]{$2.0$}
\psfrag{2.5}[c][c]{$2.5$}
\psfrag{3.0}[c][c]{$3.0$}
\psfrag{3.5}[c][c]{$3.5$}
\psfrag{Mass (keV)}[c][c]{$Mass$ (keV)}
\psfrag{CDM}[c][c]{$\rm CDM$}
\psfrag{Swift}[c][c]{\textit{Swift}}
\epsfig{trim = 0mm 4mm 4mm 16mm, clip, file=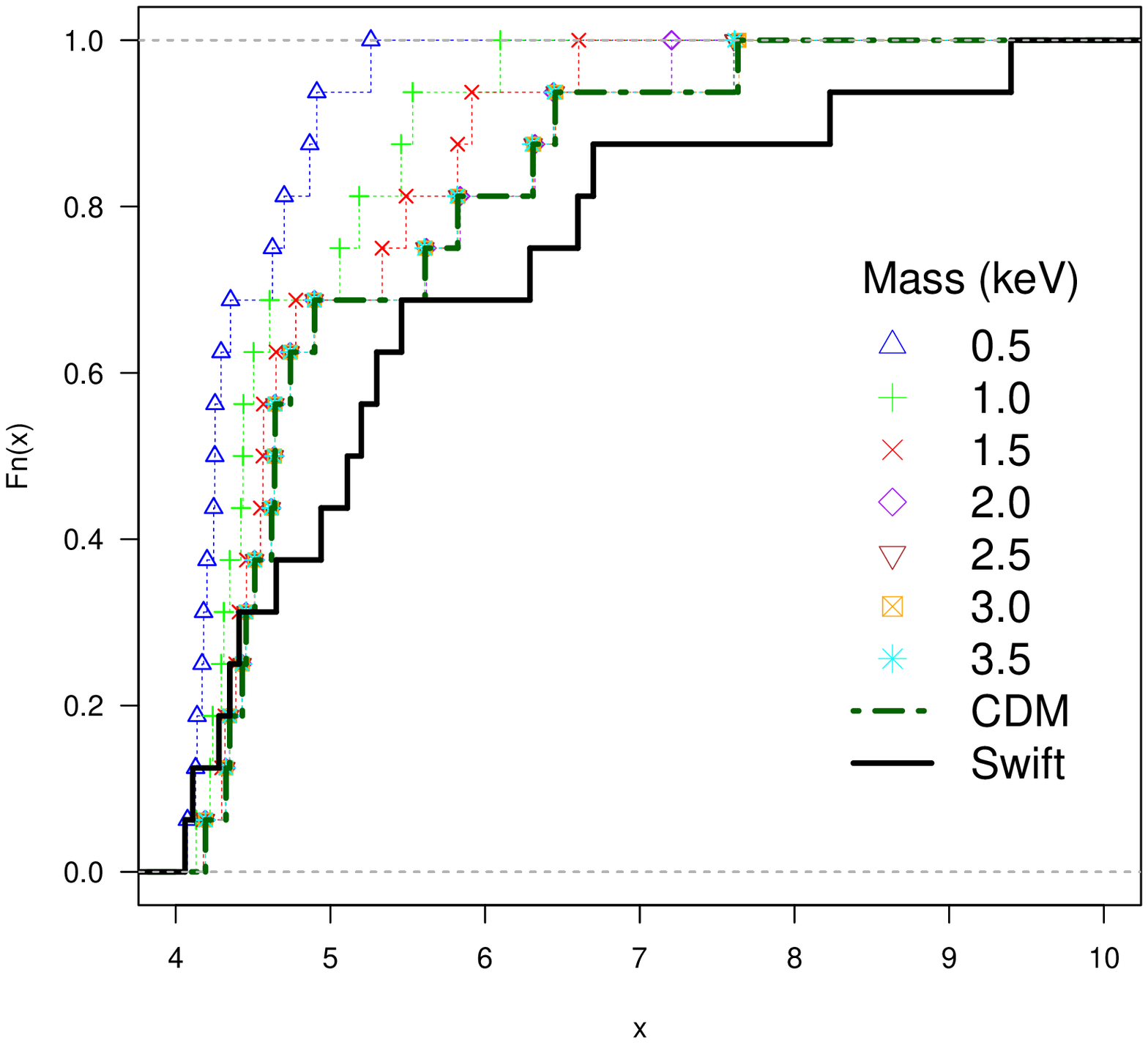,width=1\linewidth,clip=} 
\epsfig{trim = 0mm 4mm 4mm 16mm, clip, file=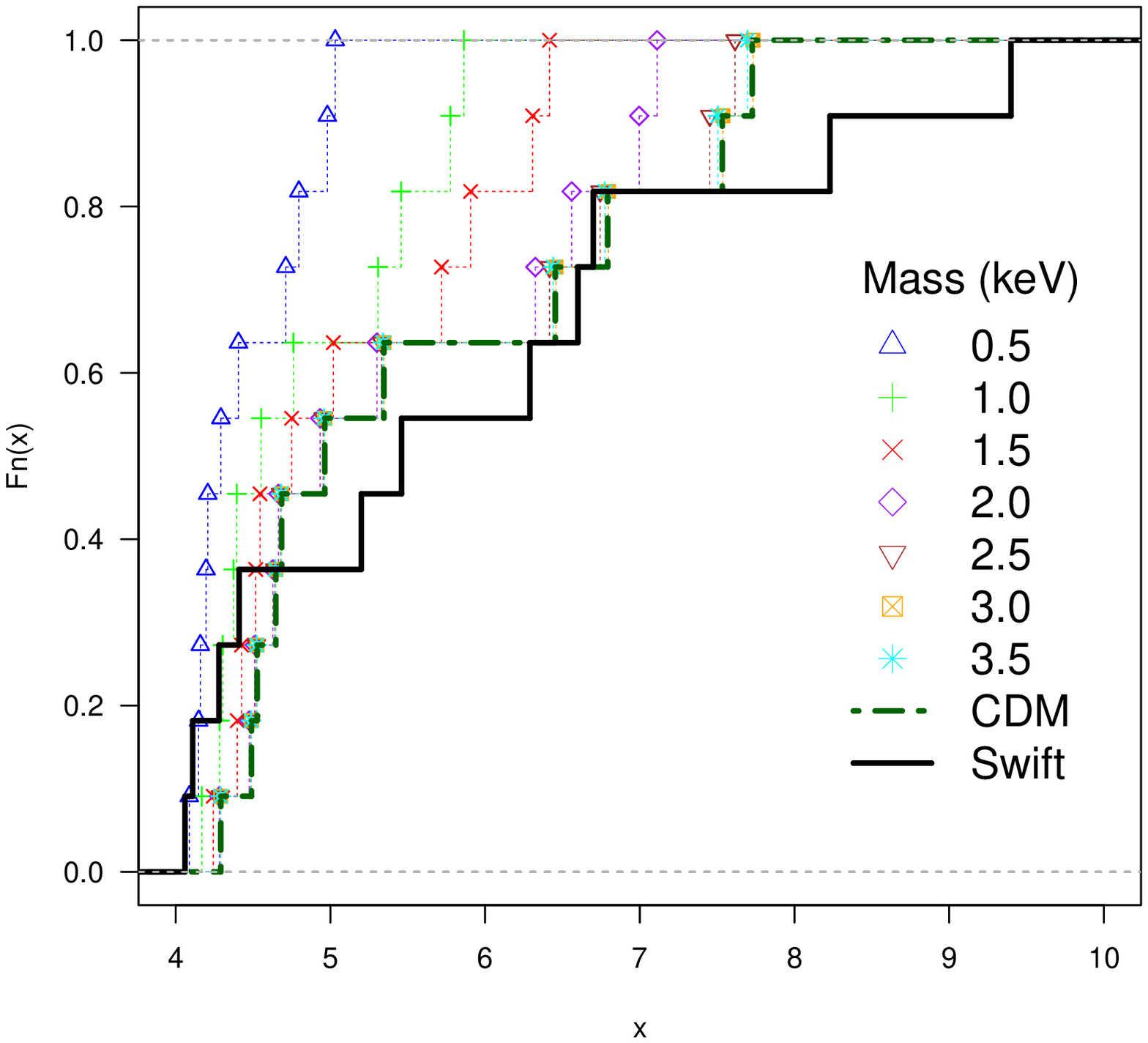,width=1\linewidth,clip=} 
\caption{Cumulative number of GRBs for different values of $m_{\rm x}$  compared with CDM predictions and $Swift$ observations. The blue dotted line corresponds to $m_{\rm x}=  0.5$  keV, green dotted line to $m_{\rm x}=  1.0$  keV, red dotted line to  $m_{\rm x}  =1.5$  keV,  purple dotted  line to  $m_{\rm x}  = 2.3$ keV, brown dotted line to  $m_{\rm x} =  2.5$  keV, orange dotted line to  $m_{\rm x} = 3.0$  keV, cyan dotted line to  $m_{\rm x}  = 3.5$ keV, dark-green two-dashed  line to  CDM, black to the \textit{Swift} observations.  \textbf{Top Panel}: sample S1C0; \textbf{Bottom panel}: sample S2C0.    }
\label{fig:ECDF}
\end{figure}

In figure \ref{fig:ECDF} we plot the CDFs for CDM and WDM (under the assumption of $\alpha=0$), as well as the observed {\it Swift} distribution.
 The lighter the WDM particle, the sharper the CDF rise at low-$z$.  There is a  
 clear separation between CDM and  WDM models with $m_{\rm x} \lesssim 1.5$ keV.  Both the S1 and S2 samples (top and bottom panels respectively) show the same qualitative trends.

\begin{figure}
\psfrag{Fn(x)}[c][c]{CDF}
\psfrag{x}[c][c]{$z$}
\psfrag{0}[c][c]{$0$}
\psfrag{3}[c][c]{$3$}
\psfrag{4}[c][c]{$4$}
\psfrag{5}[c][c]{$5$}
\psfrag{6}[c][c]{$6$}
\psfrag{7}[c][c]{$7$}
\psfrag{8}[c][c]{$8$}
\psfrag{9}[c][c]{$9$}
\psfrag{10}[c][c]{$10$}
\psfrag{12}[c][c]{$12$}
\psfrag{15}[c][c]{$15$}
\psfrag{18}[c][c]{$18$}
\psfrag{0.0}[c][c]{$0$}
\psfrag{0.2}[c][c]{$0.2$}
\psfrag{0.4}[c][c]{$0.4$}
\psfrag{0.6}[c][c]{$0.6$}
\psfrag{0.8}[c][c]{$0.8$}
\psfrag{1.0}[c][c]{$1.0$}
\psfrag{1.5}[c][c]{$1.5$}
\psfrag{2.0}[c][c]{$2.0$}
\psfrag{2.5}[c][c]{$2.5$}
\psfrag{3.0}[c][c]{$3.0$}
\psfrag{Mass (keV)}[c][c]{$Mass$ (keV)}
\psfrag{CDM}[c][c]{$\rm CDM$}
\psfrag{Swift}[c][c]{$\rm Swift$}
\epsfig{trim = 0mm 4mm 4mm 16mm, clip, file=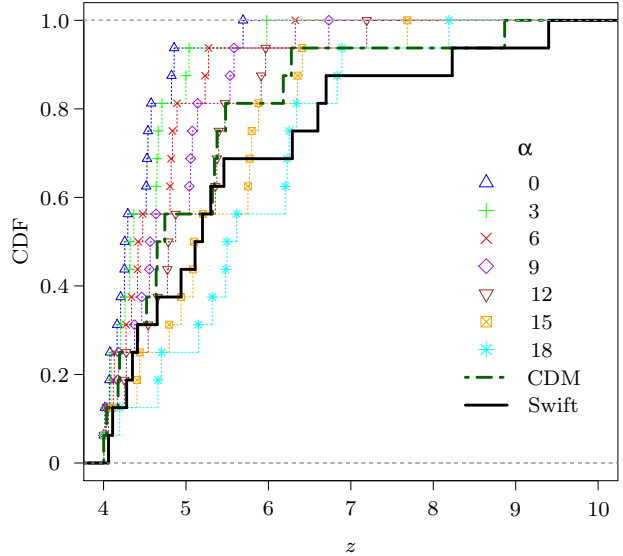,width=1\linewidth,clip=} 
\caption{Cumulative number of GRBs for $m_{\rm x} = 0.5$ keV as a function of the $\alpha$ parameter.  Blue dotted line represents  $\alpha = 0$, green dotted line $\alpha = 3$, red dotted line $\alpha = 6$, purple dotted line $\alpha = 9$,  brown dotted line $\alpha = 12$,  orange dotted line $\alpha = 15$, cyan dotted line $\alpha = 18$, dark-green two-dashed line CDM and black line the \textit{Swift} observations.  }
\label{fig:ECDF_alpha}
\end{figure}

As we saw above, the high-$z$ suppression of structures in WDM models can be compensated for by allowing the GRB rate/SFR to increase towards higher redshifts.  How degenerate are these cosmological vs astrophysical effects? In figure \ref{fig:ECDF_alpha}, we show the CDF for $m_{\rm x} = 0.5$ keV for several values of $\alpha$ for S2 sample.  The exponential suppression of DM halo abundances in this model is so strong, that an unrealistically high value of $\alpha \sim 15$ is required to  be roughly consistent with observations.  Such a high value is ruled out by low-redshift observations, which imply $\alpha\lesssim1$ (e.g. \citealt{Robertson2012, kistler2009, Trenti2012}).

\subsection{Constraints from the redshift  distribution  of $z>4$ bursts }
\label{sec:constraints}

To quantify how consistent are these CDFs with the observed distribution from {\it Swift}, we make use of two statistics: (i) the one-sample Kolmogorov-Smirnov (K-S)  test; and (ii) a maximum likelihood estimation (MLE).  Both tests are described in detail in  appendix \ref{appendix:statistics}.

The K-S test provides a simple estimate of the probability the observed distribution was drawn from the underlying theoretical one.  We compute this probability, for fixed $\alpha$ first,  for our models S1C0,  S1C1, S2C0 and S2C1.  Consistent with the more qualitative analysis from the previous section,  models with $m_{\rm x} \lesssim 1.0$ keV  are ruled out  at 90\% CL assuming $-1 \leqslant\alpha \leqslant1 $.  For  $\alpha = 0$ (S1C0 and S2C0),  the limits  are even  more restrictive  and models with $m_{\rm x} \lesssim 1.5$ keV are ruled out  at   90\% CL for both samples.

\begin{figure}
\psfrag{mx}[c][c]{$m_{\rm x}$ (keV)}
\psfrag{alpha}[c][c]{$\alpha$}
\psfrag{-2}[c][c]{$-2$}
\psfrag{2}[c][c]{$2$}
\psfrag{4}[c][c]{$4$}
\psfrag{6}[c][c]{$6$}
\psfrag{8}[c][c]{$8$}
\psfrag{0}[c][c]{$0$}
\psfrag{5}[c][c]{$5$}
\psfrag{10}[c][c]{$10$}
\psfrag{15}[c][c]{$15$}
\psfrag{20}[c][c]{$20$}
\psfrag{0.5}[c][c]{$0.5$}
\psfrag{1.0}[c][c]{$1.0$}
\psfrag{1.5}[c][c]{$1.5$}
\psfrag{2.0}[c][c]{$2.0$}
\psfrag{2.5}[c][c]{$2.5$}
\psfrag{3.0}[c][c]{$3.0$}
\psfrag{3.5}[c][c]{3.5}
\psfrag{4.0}[c][c]{CDM}

\epsfig{trim = 0mm 5mm 34mm 16mm, clip, file=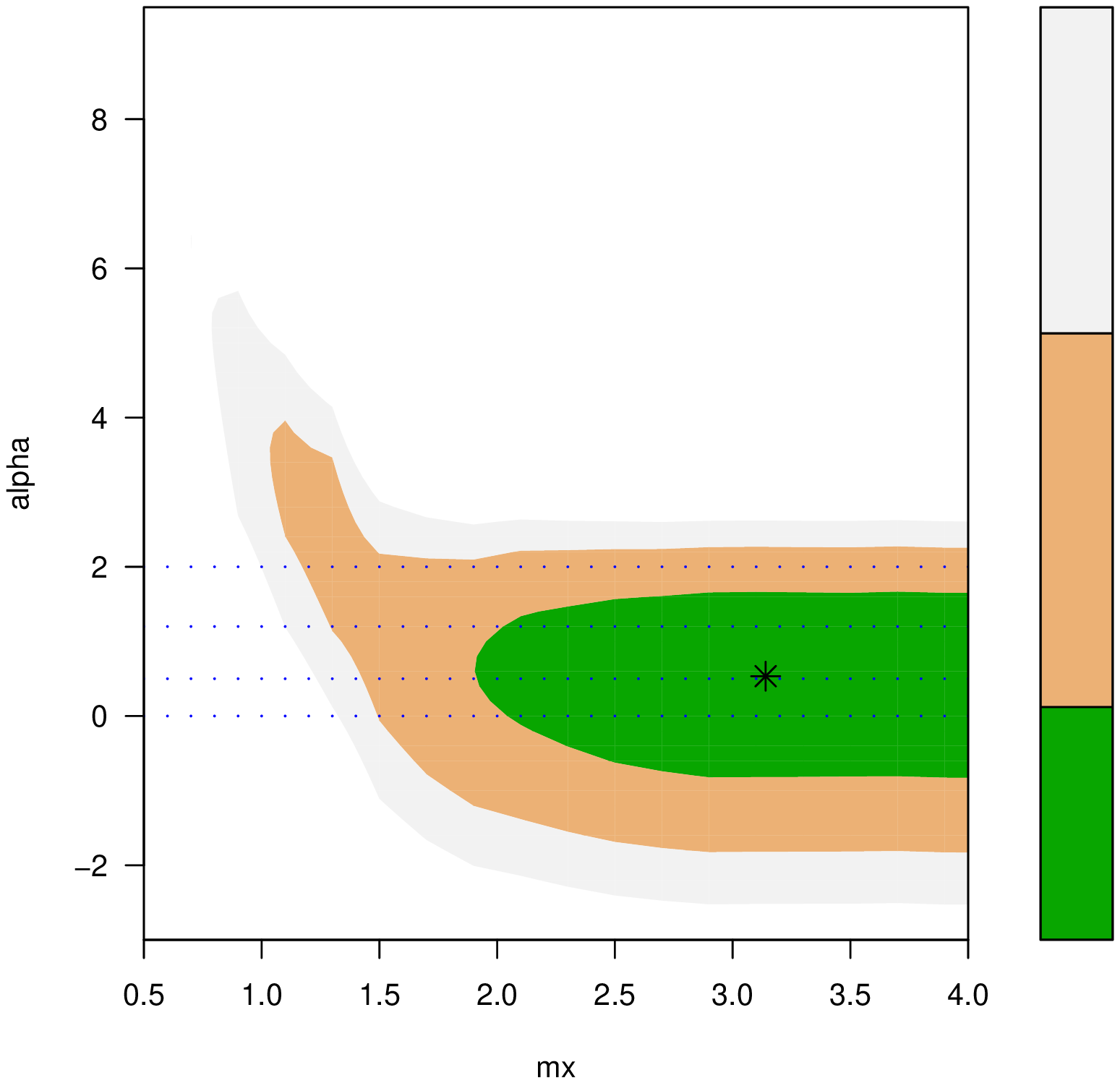,width=0.95\linewidth,clip=} 
\epsfig{trim = 0mm 5mm 34mm 16mm, clip, file=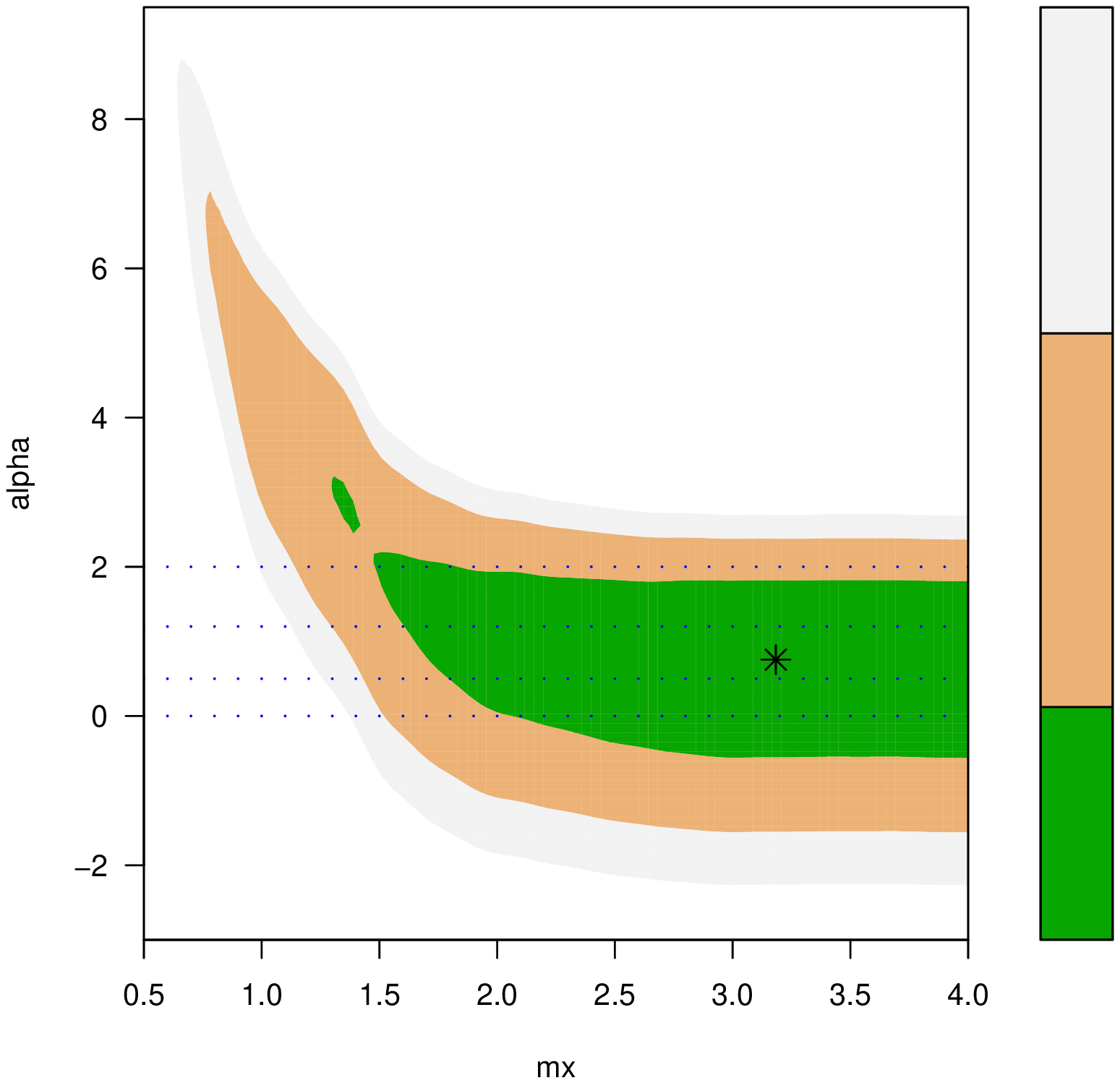,width=0.95\linewidth,clip=}

\caption{Contours over $\alpha$ and $m_{\rm x}$, enclosing 68\% (green),  95\% (orange) and 99\% (grey) probability. The asterisks correspond to the best-fit parameter combinations. \textbf{Top panel}: Sample S1C2; \textbf{bottom panel}: Sample S2C2. Horizontal  dotted  lines represent the values $\alpha = 0$ \citep{ishida2011,Elliott2012},   $\alpha = 0.5$ \citep{Robertson2012}, $\alpha = 1.2$, \citep{kistler2009} and $\alpha = 2$ for  comparison. }
\label{fig:alpha_contour1}
\end{figure}

  So far,  we have analyzed each model individually in order to quantify a lower limit on $m_{\rm x}$, given a single value of $\alpha$.  Using a $\chi^2$ MLE (see appendix \ref{appendix:statistics}) allows us to compute posterior probabilities given conservative priors on $\alpha$.  Thus we are able to construct confidence limits in the two-dimensional, ($m_{\rm x}$, $\alpha$) parameter space.
The results for cases S1C2, S2C2  are shown in figure \ref{fig:alpha_contour1} at 68\%, 95\%, 99\% CL.  Both samples show the same qualitative trends, with the data preferring higher values of $m_{\rm x}$ and CDM.   Marginalizing the likelihood  over $ -3 \leqslant \alpha \leqslant 12$,  with a flat prior,  shows that models with $m_{\rm x} \leqslant 1.6-1.8$
 keV  are ruled out at 95\% CL for S1C2 and S2C2 respectively.

\section{Future constraints }

In the previous section, we have quantified the constraints on WDM particle masses using current {\it Swift} GRB observations.  We obtain constraints of $m_{\rm x} \gtrsim 1.6$--1.8 keV.  We now ask how much could these constraints could improve with a larger GRB sample, available from future missions? 
As a reference, we use the Sino-French  space-based multi-band astronomical variable objects monitor (SVOM)\footnote{http://www.svom.fr}  mission.  The SVOM  has been designed to optimize the synergy between space and ground instruments. It is forecast to observe $\sim 70-90$ GRBs  $\rm yr^{-1}$  and $\sim 2-6$ GRB $\rm yr^{-1}$ at $z \geqslant 6$ \citep[see e.g., ][]{Salvaterra2008}.

\begin{figure}
\psfrag{mx}[c][c]{$m_{\rm x}$ (keV)}
\psfrag{alpha}[c][c]{$\alpha$}
\psfrag{-2}[c][c]{$-2$}
\psfrag{2}[c][c]{$2$}
\psfrag{4}[c][c]{$4$}
\psfrag{6}[c][c]{$6$}
\psfrag{8}[c][c]{$8$}
\psfrag{0}[c][c]{$0$}
\psfrag{5}[c][c]{$5$}
\psfrag{10}[c][c]{$10$}
\psfrag{15}[c][c]{$15$}
\psfrag{20}[c][c]{$20$}
\psfrag{0.5}[c][c]{$0.5$}
\psfrag{1.0}[c][c]{$1.0$}
\psfrag{1.5}[c][c]{$1.5$}
\psfrag{2.0}[c][c]{$2.0$}
\psfrag{2.5}[c][c]{$2.5$}
\psfrag{3.0}[c][c]{$3.0$}
\psfrag{3.5}[c][c]{3.5}
\psfrag{4.0}[c][c]{CDM}

\epsfig{trim = 0mm 5mm 34mm 16mm, clip, file=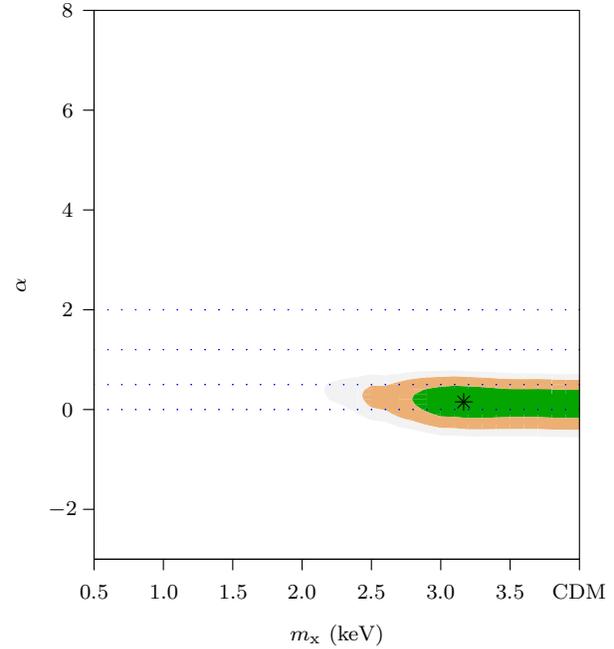,width=0.95\linewidth,clip=}  
\caption{Same as figure \ref{fig:alpha_contour1}, but assuming a 450 burst mock sample, drawn from the CDM, $\alpha=0$ PDF.}
\label{fig:alpha_contour_future}
\end{figure}

We first construct a mock GRB dataset of 450 bursts with redshifts obtained by sampling the CDM, $\alpha=0$ PDF given by equation (\ref{eq:DNDZ}).
This sample size represents an optimistic prediction   for 5 yrs of SVOM   observations\footnote{The highest redshift in our mock sample is $z_{\rm max} = 11.6$, being the only event at $z > 10$.} \citep[see e.g., ][]{Salvaterra2008}.   We then perform the MLE analysis detailed above on this mock dataset at $z>4$.  The resulting confidence limits are presented in Fig. \ref{fig:alpha_contour_future}.

 This figure shows that $\sim$ 5 yrs of SVOM observations would be sufficient to rule out $m_{\rm x} \leqslant 2.3$ keV models (from our fiducial CDM, $\alpha=0$ model) at 95\% CL, when marginalized   over $\alpha$. This is a modest improvement over our current constraints using {\it Swift} observations.  As already foreshadowed by figures 1 and 2, as well as the associated discussion, it is increasingly difficult to push constraints beyond $m_{\rm x}>2$ keV.  On the other hand, the $\alpha$ constraint improves dramatically due to  having enough high-$z$ bursts to beat the Poisson errors.   
 We caution that the relative narrowness around $\alpha=0$ of the contours in Fig. \ref{fig:alpha_contour_future} is also partially due to our choice of (CDM, $\alpha=0$) as the template for the mock observation.

\section{Conclusion}
\label{sec_disc}

Small-scale structures are strongly suppressed in WDM cosmologies.  WDM particle masses of $m_{\rm x}\sim$ keV have been invoked in order to interpret observations of local dwarf galaxies and galactic cores.  The high-redshift Universe is a powerful testbed for these cosmologies, since the mere presence of collapsed structures can set strong  lower limits on $m_{\rm x}$. GRBs, being extremely bright and observable to well within the first billion years, are a promising tool for such studies.

Here we model the collapsed fraction and cosmic SFR in CDM and WDM cosmologies, taking into account the effects of both free-streaming and effective pressure due to the residual velocity dispersion of WDM particles. Assuming that the GRB rate is proportional to the SFR, we interpret 5 years of {\it Swift} observations in order to place constraints on $m_{\rm x}$.  We conservatively account for astrophysical uncertainty by allowing the GRB rate/SFR to evolve with redshift as $\propto (1+z)^{\alpha}$.  In order to fold completeness limits into our analysis, we used a low-z sample to estimate the intrinsic LF, or else restricted our analysis to a luminosity-limited subsample detectable at all redshifts.

For each model ($m_{\rm x}, \alpha$), we compute both the absolute detection rates and CDFs, at $z>4$.  A K-S test between the model and observed CDFs rules out $m_{\rm x} < 1.5$ (1.0) keV, assuming $\alpha = 0$ ($<2$), at 90\% CL.
  Using a maximum likelihood estimator, we are able to marginalize over $\alpha$.  Assuming a flat prior in $\alpha$, we constrain $m_{\rm x} > 1.6$--1.8 keV at 95\% CL.  A future SVOM-like  mission would tighten these constraints to $m_{\rm x} \gtrsim 2.3 $ keV.

The strong and robust constraints we derive show that GRBs are a powerful probe of the early Universe.
Their utility would be further enhanced with insights into their formation environments and their relation to the cosmic SFR.

\section*{Acknowledgements}

We thank Emille Ishida   for the  careful and fruitful  revision of the draft of this work  and Andressa Jendreieck for useful  comments. RSS thanks the Max Planck Institute for Astrophysics (Garching, Germany) for its hospitality during his visit.

\bsp

\appendix
\section{Statistics}
\label{appendix:statistics}

 \subsubsection*{Kolmogorov-Smirnov one-sample test}
 A straightforward way to compare the  data and WDM models is to  perform  a one-sample Kolmogorov-Smirnov (K-S) test.  The null hypothesis that the observed GRB redshifts  are consistent with a model distribution can be evaluated by estimating  a  \textit{p-value}, which  corresponds to one minus the probability that the null hypothesis can be rejected. The K-S test consists in comparing the statistical parameter  
 \begin{equation}
 D = \mathrm{sup}|F(z)-G(z)|,
 \end{equation}
 where $F(z)$ and $G(z)$ are the CDF for the theoretical and observed sample and $\mathrm{sup}$ is the  the supremum of a  totally or partially ordered set.  We estimate the  \textit{p-value} via nonparametric bootstrap, which consists of running   Monte Carlo realizations of the observed CDF  using a random-selection-with-replacement procedure estimated from the data. This provides a   histogram of the statistic D, from which a valid goodness-of-fit probability can be evaluated. The  probability distribution function for each model is determined   by  equation (\ref{eq:Nz1z2}). 

 \subsubsection*{Maximum-likelihood method}
More formally, we can estimate the probability of parameters $\{\alpha,m_{\rm x}\}$ given the observed data  using a Bayesian technique. Assuming that our data is described by the probability density function $f(x;\pmb{\theta})$, where $x$ is a variable and $\pmb{\theta} \equiv \{\alpha,m_{\rm x}\}$.   We want to estimate $\pmb{\theta}$, assuming the data are independent.  So the likelihood will be given by 
\begin{equation}
{\cal L}(\pmb{\theta}|x_i)\propto  \prod_{i=1}^{N}f(x_i|\pmb{\theta}).
\end{equation}

Given the small number of observed bursts  per  redshift  bin, $\Delta z = 1.5$, we use a  Poisson error statistics\footnote{The value $\Delta z = 1.5$ is chosen to ensure at least 1 burst per bin.} (see e.g., \citealt{Campanelli2012} for a similar procedure applied to galaxy cluster number count). Therefore, the likelihood function can be computed as  

\begin{equation}
{\cal L}(\pmb{\theta}|\kappa_i)\propto  \prod_{i=1}^{5}\frac{\Upsilon_i^{\kappa_i}e^{-\Upsilon_i}}{\kappa_i!}, 
\end{equation} 
where  $\Upsilon_i \equiv N(z_i,z_{i+1};\pmb{\theta})$ and ${\kappa_i} \equiv  N_{\rm obs}(z_i,z_{i+1})$. Thus, the $\chi^2$ statistics can be written as 
\begin{eqnarray}
\chi^2(\pmb{\theta}) &=& -2\ln{\cal L},\nonumber\\
&=& 2\sum_{i=1}^{5}\Upsilon_i-\kappa_i(1+\ln{\Upsilon_i}-\ln{\kappa_i)}.
\end{eqnarray}

\label{lastpage}
\end{document}